\documentclass[usenatbib,useAMS]{mn2e}
\usepackage{epsfig}
\usepackage{times}
\usepackage{amsmath}

\usepackage{verbatim}

\title[Light curves for off-centre SN~Ia explosions]{Light curves for off-centre ignition models of type Ia supernovae}
\author[Sim et al.]{S. A. Sim$^1$, D. N. Sauer$^1$, F. K. R\"opke$^{1,2}$, W. Hillebrandt$^1$\\
$^{1}$Max-Planck-Institut f\"{u}r Astrophysik, Karl-Schwarzschildstr. 1,
85748 Garching, Germany\\
$^{2}$Department of Astronomy and Astrophysics, University of
California, Santa Cruz, 1156 High Street, Santa Cruz, CA 95064, USA
}

\date{\today}

\pagerange{\pageref{firstpage}--\pageref{lastpage}}
\pubyear{}
\begin{document}
\maketitle
\label{firstpage}

\begin{abstract}
Motivated by recent models involving off-centre ignition of type Ia supernova
explosions, we undertake three-dimensional time-dependent radiation transport
simulations to investigate the range of bolometric light curve properties that
could be observed from supernovae in which there is a lop-sided
distribution of the products from nuclear burning. We consider both a grid of
artificial toy models which illustrate the conceivable range of effects and a
recent three-dimensional hydrodynamical explosion model.  We find that
observationally significant viewing angle effects are likely to arise in such
supernovae and that these may have important ramifications for the
interpretation of the observed diversity of type Ia supernova and the
systematic uncertainties which relate to their use as standard candles in
contemporary cosmology.
\end{abstract}


\section{Introduction}

The most popular progenitor model for the average type Ia supernova
(SN~Ia) is a massive white dwarf, consisting of carbon and oxygen,
which approaches the Chandrasekhar mass by an as yet unknown mechanism,
presumably accretion from a companion star, and is disrupted by a
thermonuclear explosion (see, e.g., \citealt{hillebrandt00} for a
review).  This general picture is supported by the similarity of their
photometric and spectroscopic properties which, after calibrating
their peak luminosity by means of distance-independent properties,
turns them into very good distance indicators for cosmology
(\citealt{riess98}; \citealt{perlmutter99}). While, however, the number of 
supernovae observed at cosmological redshifts is steadily increasing
(\citealt{tonry03}; \citealt{riess04}; \citealt{astier06}; \citealt{clocchiatti06}), thus reducing
the statistical errors of the cosmological parameters derived from
them, a sound theoretical understanding of these objects -- justifying
in particular the calibration techniques applied for distance
measurements -- is still lacking.

Advances in numerical methods and the constant increase of computational power
have allowed the development of three-dimensional (3D) simulations of the
explosion phase of SNe~Ia (\citealt{reinecke02}; \citealt{gamezo03};
\citealt{roepke05}; \citealt{roepke06}). These facilitate a
consistent treatment of instabilities and turbulence effects which play a key
role in the explosion mechanism. This level of sophistication allows the models
to gain a high predictive power. At the same time, they provide the tools to
address asymmetry effects in the explosion phase. The full 3D
information from simulations has so far only been used to derive nebular
spectra \citep{kozma05}, while synthetic light curves were obtained from
spherically averaged results of such simulations \citep{blinnikov06}.

As the flame starts out in the sub-sonic deflagration burning mode, it is
subject to buoyancy instabilities leading to complex morphologies of the ash
region. While a turbulent cascade interacts with the flame on smaller scales,
on the largest scales burning bubbles float towards the star's surface evolving
into a mushroom-like shape and partially merging with each other. The first
full-star SN~Ia simulation \citep{roepke05} demonstrated that this process does
not lead to global asymmetries by favouring low-order moments in the flow.
Therefore, although complex in structure, the ash of the deflagration phase may
be volume-filling on average if ignited in an initial flame structure that was
uniformly distributed around the centre \citep{roepke05}. The flame ignition,
however, is an uncertain process and it is possible that it proceeds in a very
asymmetric manner \citep{kuhlen06, hoeflich02}. The consequence thereof may be
an asymmetric flame evolution and the most extreme example is a flame ignited
on only one side of the star. Floating rapidly towards the star's surface due
to buoyancy, it may consume only material of one side of the star
\citep{zingale_dursi}. In some cases such models fail to release sufficient
amounts of nuclear energy to overcome the gravitational binding of the star
\citep{calder04, plewa04}. Other cases, however, may explode the white dwarf and
lead to very asymmetric compositions of the ejecta. Such a model will be
discussed in Section~\ref{sect:3d}.

Another possible source of asymmetries in the ejecta is the propagation of a
detonation wave.  In delayed detonation models, such a wave is hypothesised to
trigger after a period of burning in the deflagration phase
\citep{khokhlov91}.
As recently discussed by \citet{mazzali07}, this class of model is particularly 
appealing since it provides a scenario which could be able to explain 
most SNe~Ia. 
One possibility for a deflagration-to-detonation transition (DDT) is the onset
of the distributed burning regime \citep{niemeyer_woosley}, when turbulence
first penetrates the internal flame structure. 
This happens at low fuel
densities and consequently the DDT may take place at the outer edges of the
deflagration structure. The detonation propagating inwards from this spot may
not catch up with the rapidly expanding structures on the far side before the
fuel densities fall below the burning threshold; in this case, an asymmetric
composition of the ejecta is expected and corresponding 3D
simulations have been presented by \citet{roepke07b}. It is possible, however,
that the DDT happens at multiple locations washing out the asymmetries in the
ejecta. In an alternative scenario, the ``Gravitationally Confined
Detonation'' (GCD) suggested by \citet{plewa04}, an
asymmetrically ignited flame fails to unbind the star and the ash erupts from
the surface. Still gravitationally bound, it sweeps around the unburned core of
the white dwarf and collides on the far side. The resulting 
compression of fuel 
has been suggested to trigger a detonation \citep{plewa04} which
potentially produces asymmetric ejecta compositions. However, recent
results by \citet{roepke07} indicate that triggering a detonation in
this scenario is possible only with special ignition setups in
two-dimensional simulations. Three-dimensional models seem to
disfavour this mechanism, because in those models greater energy  
release in the
deflagration stage expands the star such that a strong collision of
the ashes is prevented. In some cases the energy release during
deflagration is even sufficient to
unbind the white dwarf (an example of one such model is discussed in
Section~\ref{sect:real_model} of this paper).

The asymmetries produced in delayed detonation scenarios merit further
investigation and will be the subject of a forthcoming study. Here, we are
motivated to investigate and illustrate the possible effects of chemical
asymmetries on bolometric light curves. To this end, we will consider first a
set of toy models which explore the effect of an off-centre
distribution of nuclear ash. We then extend the discussion to a
real hydrodynamical explosion model -- an asymmetrically ignited pure
deflagration model that produced a weak explosion.

The radiative transfer calculations required to obtain the light curves were
performed using a 3D, fully time-dependent Monte Carlo code. 
We begin, in Section~\ref{sect:code}, by briefly describing the
operation of this code.
Then, in
Section~\ref{sect:toy}, we discuss our set of artificial toy
models. 
In Section~\ref{sect:real_model}, we present the results
obtained with a recent hydrodynamical explosion model \citep{roepke07} and in
Section~\ref{sect:implic} we discuss the implications of this model. Our
findings are summarised in Section~\ref{sect:summ}.

\section{Radiative transfer calculations}
\label{sect:code}

Monte Carlo
methods have been successfully applied to the modelling of radiation
transport in supernovae for more than two decades.
\citet{ambwani88}, and several subsequent studies, 
have used Monte Carlo 
simulations of $\gamma$-ray propagation to compute energy deposition
rates and $\gamma$-ray spectra as functions of time for SN~Ia.
Subsequently, the Monte Carlo approach was extended to follow not only
the $\gamma$-rays but also the subsequent emission and scattering of
radiation in other spectral regions, thereby allowing bolometric light
curves to be obtained \citep{cappellaro97}.
Recently, \citet{lucy05} has presented further generalisation of the
methods and demonstrated that they can be readily used
for three dimensional modelling. Following \citet{lucy05}, these
methods have been employed in
a variety of contemporary multi-dimensional radiative transfer
computations of relevance to the study of SNe \citep[see,
e.g.,][]{kasen06a, maeda06}.

All the light curve calculations presented in this paper
were performed using the 3D Monte
Carlo radiative transfer code described by \citet{sim07}, which is
closely based on that presented by \citet{lucy05}. A brief description
of the operation of the code is given below but we refer the reader to
\citet{sim07} for full details.

The code assumes that the ejecta are in homologous expansion, an
excellent approximation for the entire time span for which any
significant radiative flux is able to propagate and escape. The
supernova model is specified on a 3D
Cartesian grid which expands smoothly along with the homologous
flow. For the toy models (Section~\ref{sect:toy}) a 100$^3$ grid is
adopted while for the full explosion model
(Section~\ref{sect:real_model}) a 128$^3$ grid is used.
The model specifies the initial density in each grid cell, 
the initial fractional mass of $^{56}$Ni and, when needed, the
combined fractional
masses of all iron group elements. The mass density at later times is
readily deduced from the assumption of homologous expansion.

Following \citet{lucy05}, the Monte Carlo quanta begin their lives as
pellets of radioactive material. These pellets are placed in the
ejecta in accordance with the initial distribution of $^{56}$Ni. As
time passes in the simulation, the pellets decay as determined by the
half-lives of the $^{56}$Ni and daughter $^{56}$Co nuclei. When a
pellet decays, it becomes a $\gamma$-ray quantum, the propagation of
which is then followed in detail. By means of Compton scattering or
photoabsorption, the $\gamma$-ray quanta are able to deposit their
energy -- the numerical treatment of these processes is discussed by 
\citet{lucy05}. Whenever a $\gamma$-ray quantum is destroyed, it is
assumed that its energy thermalises and is re-emitted as 
ultraviolet, optical or infrared ({\sc uvoir}) photons; it is
thereafter termed an $r$-packet in the nomenclature used
by \citet{lucy05}.

Bolometric {\sc uvoir} light curves are
obtained from the behaviour of the $r$-packets. Currently, the code
uses a simple
grey-opacity treatment to follow the propagation and
scattering of the $r$-packets (the specific forms of the opacity used
for the models we present are discussed in Sections~\ref{sect:toy} and 
\ref{sect:real_model}). In principle, light curves can be obtained by
angular binning of the $r$-packets as they emerge from the computational
domain. However, greater efficiency is obtained by employing
volume-based 
Monte Carlo estimators \citep{lucy99} to deduce emissivities from the
$r$-packet trajectories and then using these to perform a formal
solution of the radiative transfer equation (details of the
specific Monte Carlo estimators currently used in the code are given 
in \citealt{sim07}). Thus, the end products of the calculation which
are used in this paper are
bolometric light curves computed for distant observers who are located
on specific lines-of-sight and who are at rest relative to the centre
of the supernova model.

\begin{figure*}
\epsfig{file=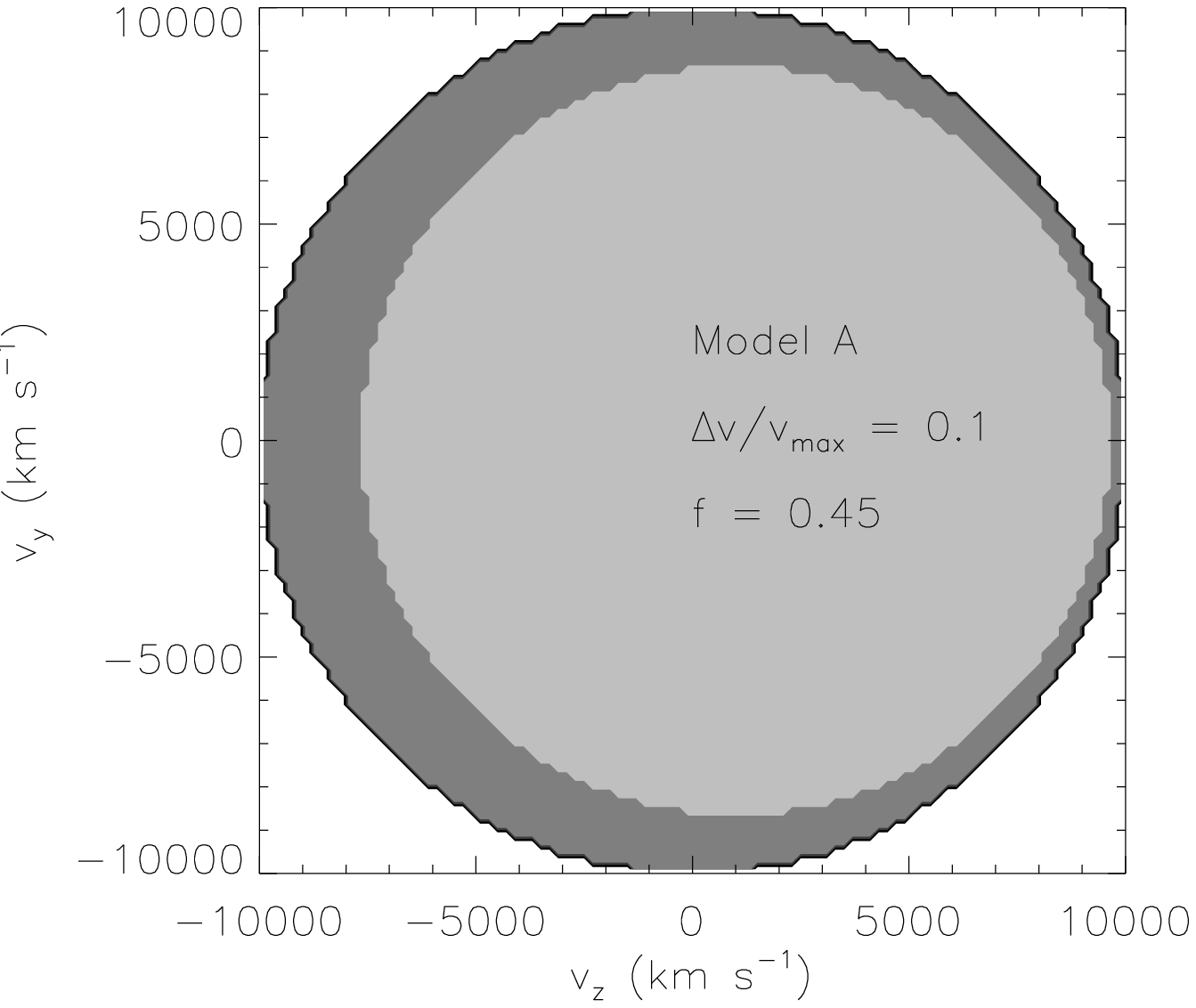, width=\columnwidth}
\epsfig{file=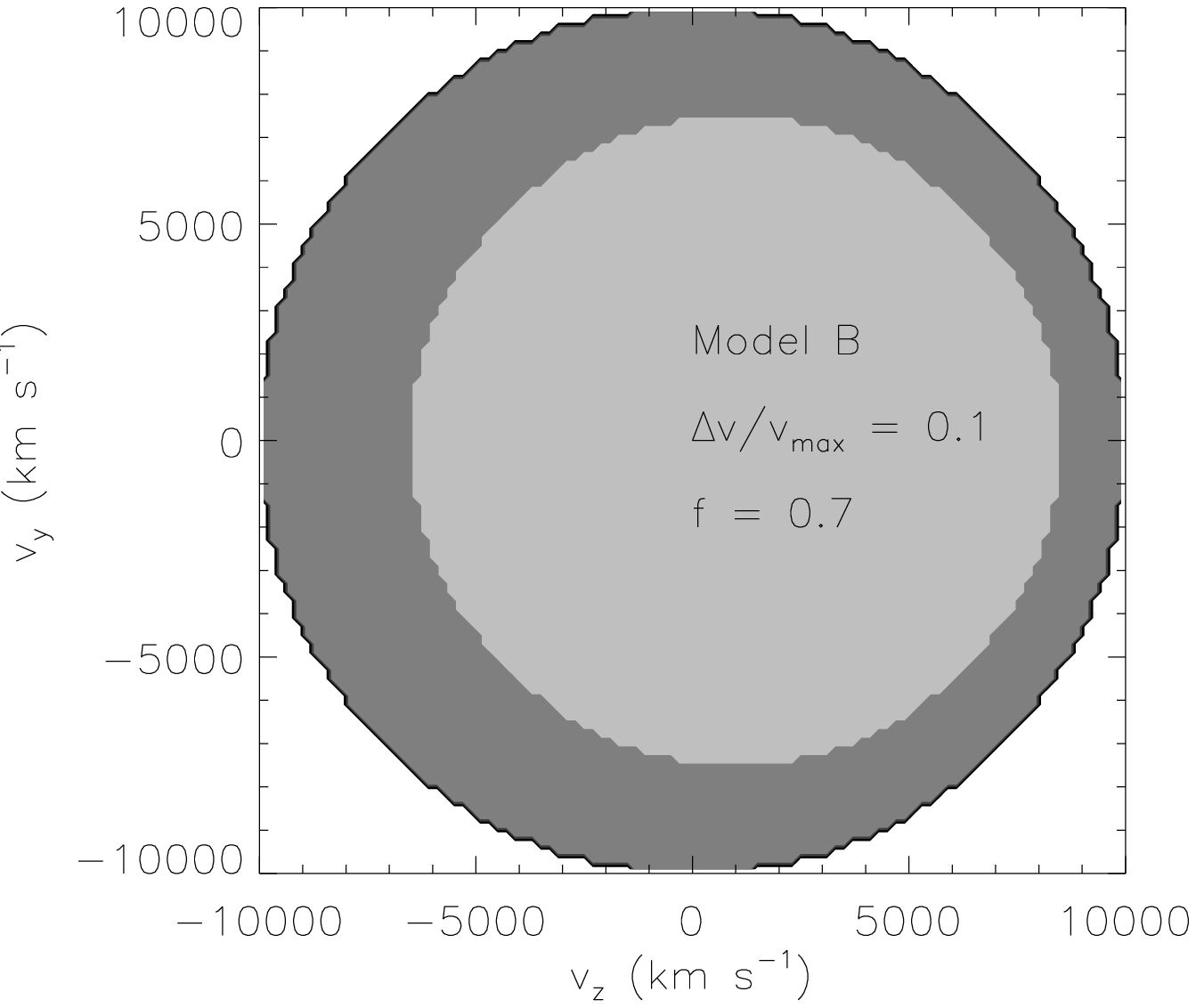, width=\columnwidth}\\
\epsfig{file=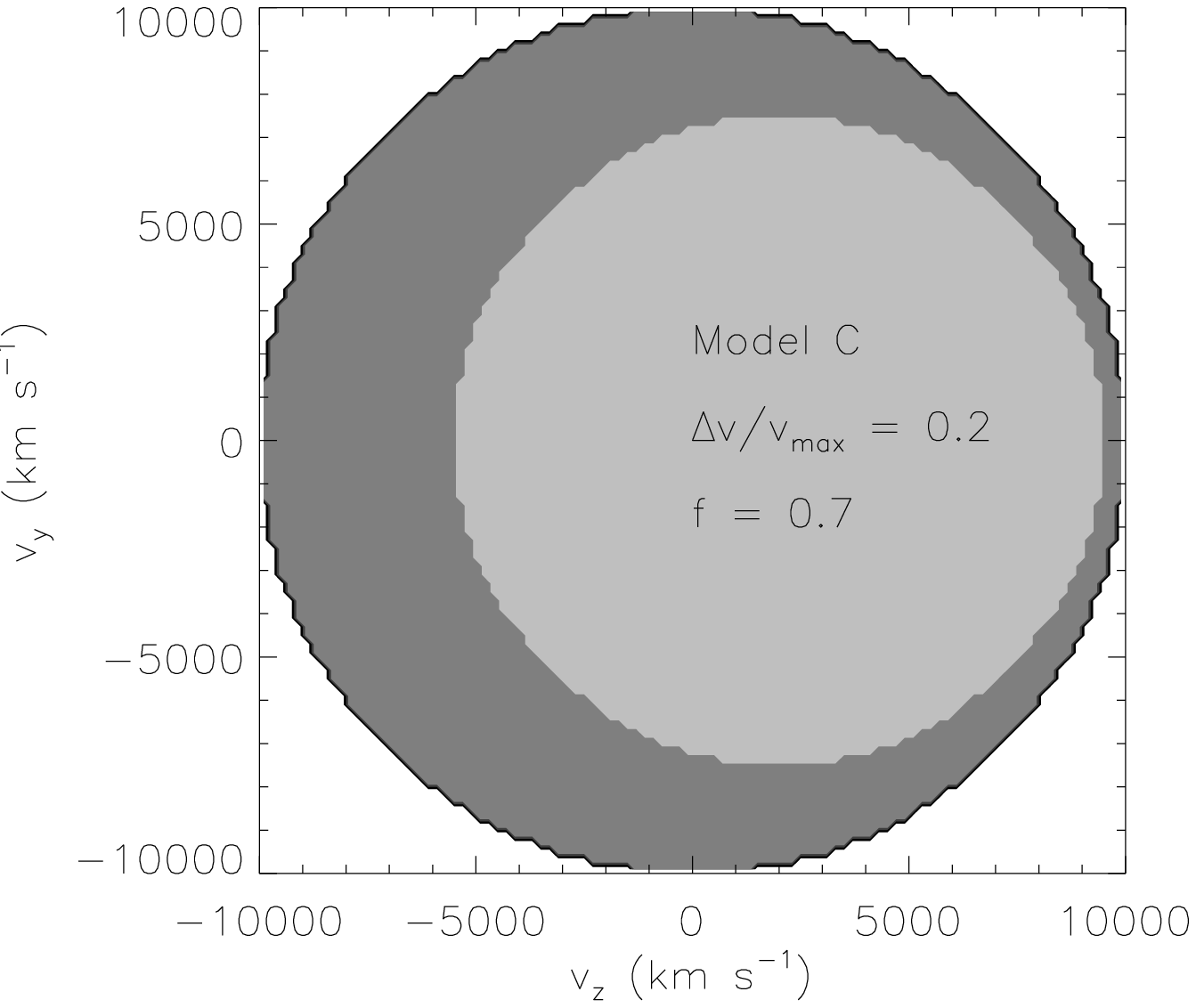, width=\columnwidth}
\epsfig{file=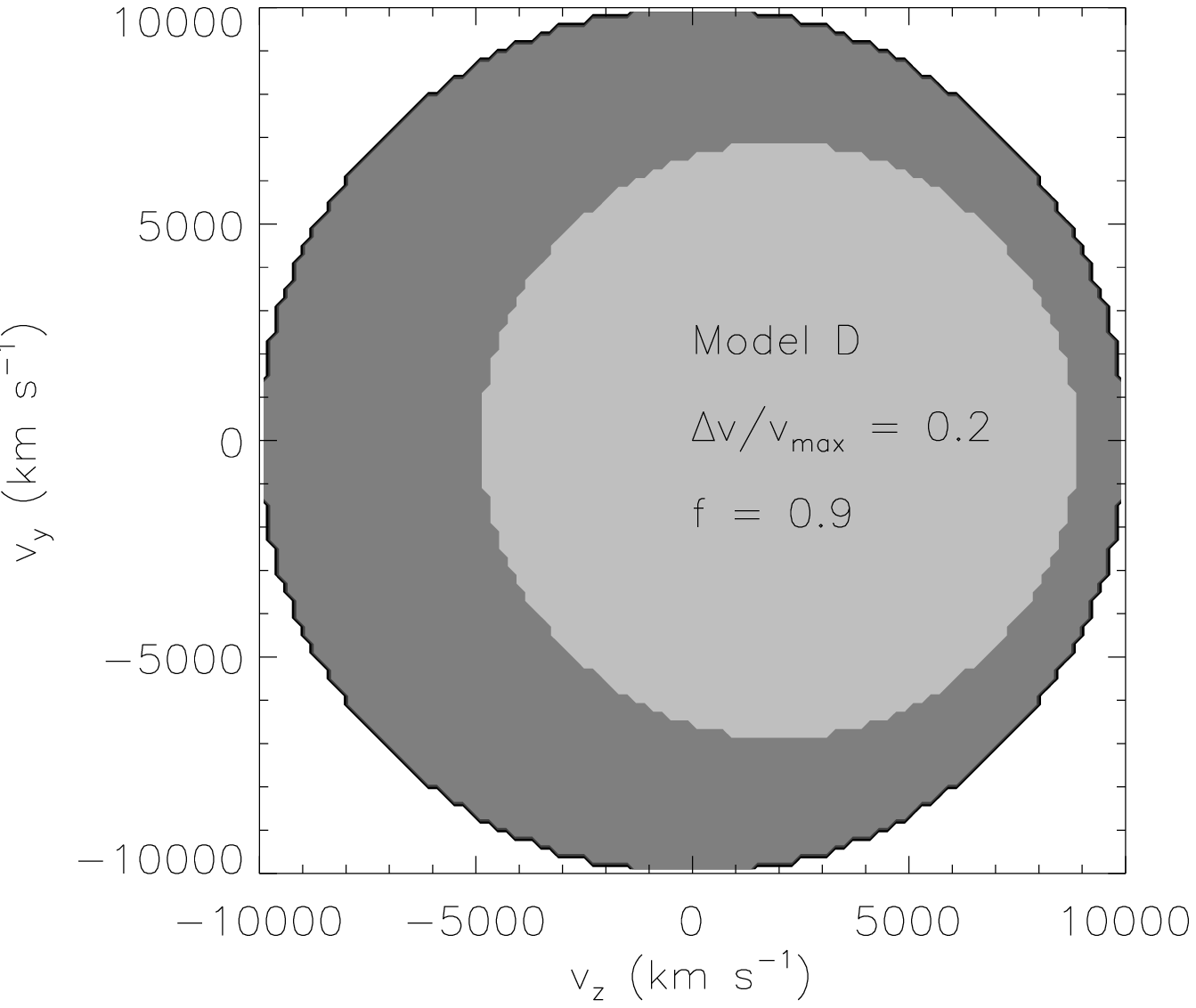, width=\columnwidth}\\
\epsfig{file=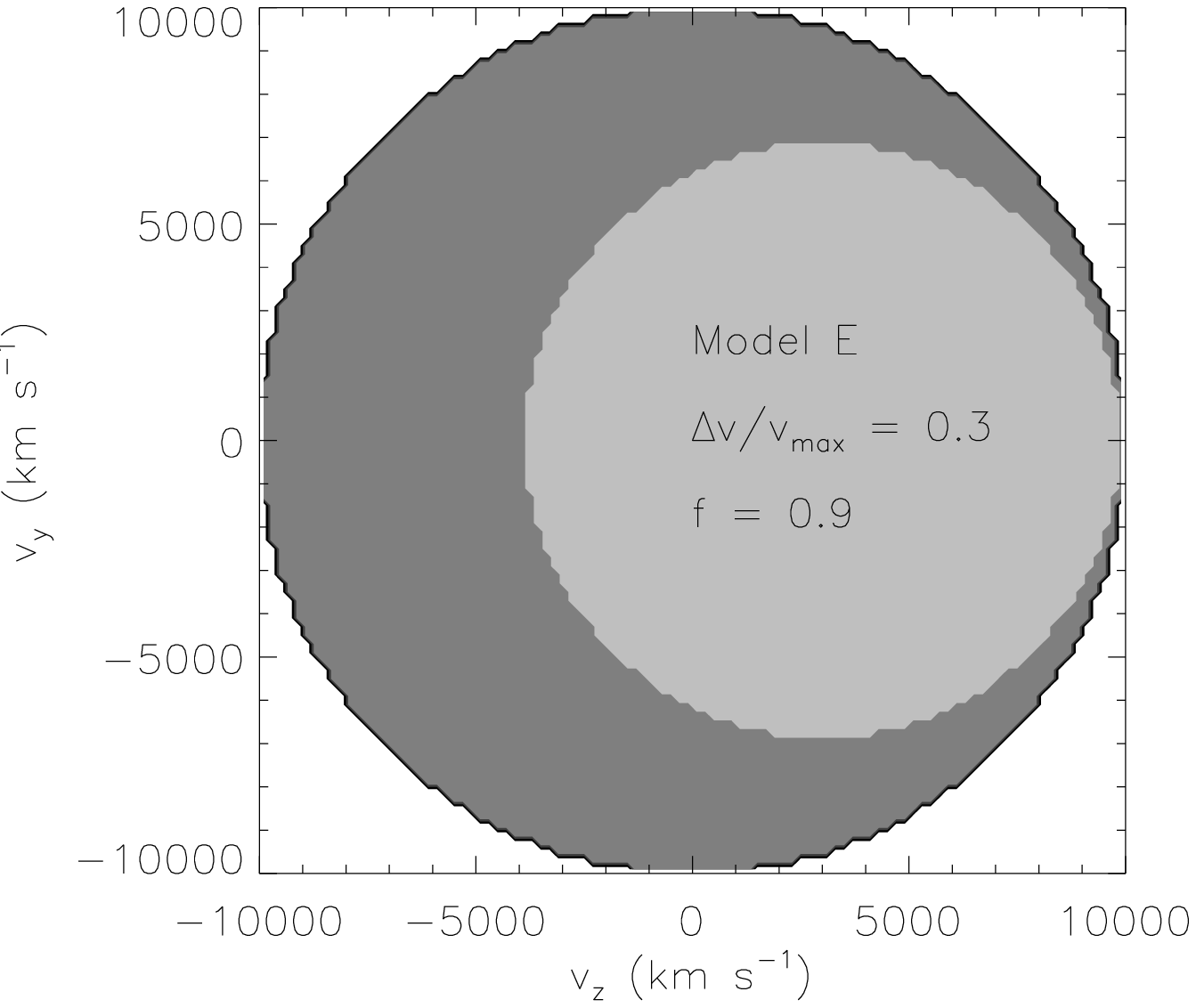, width=\columnwidth}\\
\caption{Slices through the y-z plane showing the distribution of
$^{56}$Ni adopted in the models. The dark grey areas indicate the
regions in which there is no $^{56}$Ni while the light grey areas are
those containing a fractional mass $f$ of $^{56}$Ni. The adopted
values of $f$ are indicated in the figure. $\Delta v /
v_{\mbox{\scriptsize max}}$ specifies the fractional offset in
velocity of the $^{56}$Ni centre-of-mass from the true centre-of-mass
of the SN. All the models have a total mass of 1.4~M$_{\odot}$ and a
total $^{56}$Ni mass of 0.4~M$_{\odot}$. All the models are symmetric
under rotation about the $z$-axis.
\label{fig:toy-pics}
}
\end{figure*}

\section{Toy models}
\label{sect:toy}

In this section, we undertake a preliminary investigation of the effects of a
lop-sided $^{56}$Ni distribution on light curves using a set of
simply-parameterized, artificial toy models. We begin by describing these
models and the numerical results obtained from them (Section
\ref{sect:toy-models}) and then discuss the general implications of such models
(Section \ref{sect:toy-discuss}).

\subsection{Description}
\label{sect:toy-models}

To explore the role of an off-centre distribution of $^{56}$Ni, a grid of
simplistic toy models has been constructed. In each model, a total
mass of 1.4~M$_{\odot}$ is adopted.
The mass density is assumed to be uniform and a 
maximum velocity of $v_{\mbox{\scriptsize max}} = 10^4$~km~s$^{-1}$
is chosen.
This simple density distribution makes it convenient to explore
off-centre distributions of $^{56}$Ni without necessitating changes in
either the geometrical shape of the $^{56}$Ni blob nor the underlying
total mass distributions.
In more sophisticated SN~Ia models, the density distribution generally
does extend to velocities greater than our adopted 
$v_{\mbox{\scriptsize max}}$ (e.g. in the well-known W7 model, 
\citealt{nomoto84}) -- however, in these outer higher velocity regions,
the density has typically fallen off by an order of magnitude or more
compared to the inner parts which makes the outer regions relatively
unimportant in grey-bolometric light curve calculations such as
employed here. In particular, \citet{pinto00a} demonstrated that the
light curve computed from a uniform-density spherical model with 
$v_{\mbox{\scriptsize max}} = 10^4$~km~s$^{-1}$ agrees very well with
that obtained from the W7 model (see their figure~2).

Inside our toy SN models, the $^{56}$Ni is also
given a spherical distribution but its centre-of-mass is offset in velocity
relative to that of the SN; this offset ($\Delta v$, which will generally be
specified as a fraction of $v_{\mbox{\scriptsize max}}$) is one of the
parameters of the models. Throughout the region which contains $^{56}$Ni, a
uniform mass-fraction $f$ is adopted -- this is the second
parameter of the model. In all the models the total mass of
$^{56}$Ni is fixed to 0.4~M$_{\odot}$; therefore $f$ determines 
the volume of the region which contains
$^{56}$Ni.  In total, five models (A -- E) are presented here; the structure of
these models is illustrated in Figure~\ref{fig:toy-pics}. The parameters which
differentiate 
the models ($\Delta v/v_{\mbox{\scriptsize max}}$, $f$) are indicated in
the figure and also tabulated in Table~\ref{tab:toy}.  In all the toy models, a
uniform grey-absorption cross section of 0.1~cm$^{2}$~g$^{-1}$ is
adopted for the treatment of ``bolometric''
{\sc uvoir} photons while the
$\gamma$-ray transport and deposition is treated in detail (see
\citealt{sim07} for a full description).

Light curves have been computed for each of the five models (A -- E) for three
interesting viewing directions: 

\begin{enumerate}
  \item for an observer located at a large distance along the same direction in
    which the Ni is displaced relative to the centre of the SN; 
  \item for an observer located diametrically opposite the observer in case (i); 
  \item for an observer lying along a line-of-sight perpendicular to the
    direction of displacement. 
\end{enumerate}
For each of these viewing angles in each of the models, the time ($t_{\rm p}$) and
magnitude ($M_{\rm p}$) of maximum light have been extracted; these values are
reported in Table~\ref{tab:toy}. 

The results in Table~\ref{tab:toy} indicate that an off-centre distribution of
$^{56}$Ni could introduce significant angular dependence in the light curve
properties.  In
the models, the light curve peak is both earliest and brightest when viewed
along the direction in which the $^{56}$Ni is displaced. This behaviour is as
expected since the mean optical depth from the $^{56}$Ni to the edge of the
ejecta is lowest in this direction; therefore, more energy packets escape more
quickly.  Correspondingly, when viewed from the diametrically
opposite direction, the peak magnitude occurs latest and is dimmest.

It can be seen that the viewing-angle sensitivity of $M_{\rm p}$ is significantly
affected by $\Delta v$ but is insensitive to the adopted value of $f$. As
$\Delta v$ is increased, the variation of $M_{\rm p}$ with angle in a given model
increases from about 0.5~mag for the models with
$\Delta v / v_{\mbox{\scriptsize max}} = 0.1$ to more than 1.5~mag for the most
extreme model considered ($\Delta v / v_{\mbox{\scriptsize max}} = 0.3$).

\begin{figure}
\epsfig{file=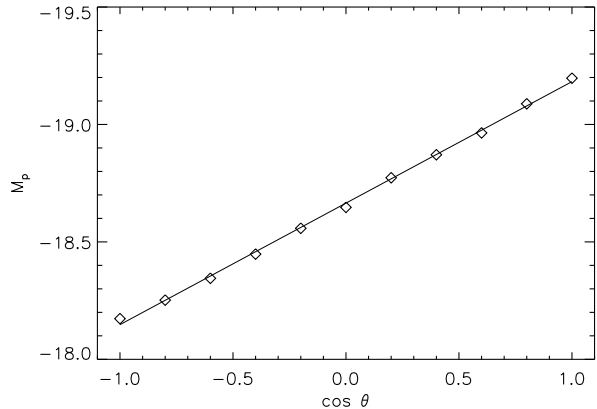, width=\columnwidth}
\caption{The peak bolometric magnitude ($M_{\mbox{\scriptsize p}}$) as a
  function of the viewing direction ($\theta$) for light curves computed with
  model~C (shown as diamonds). The solid line is the linear fit to the results
  (see text).  \label{fig:toy-mpeak} }
\end{figure}

\begin{figure}
\epsfig{file=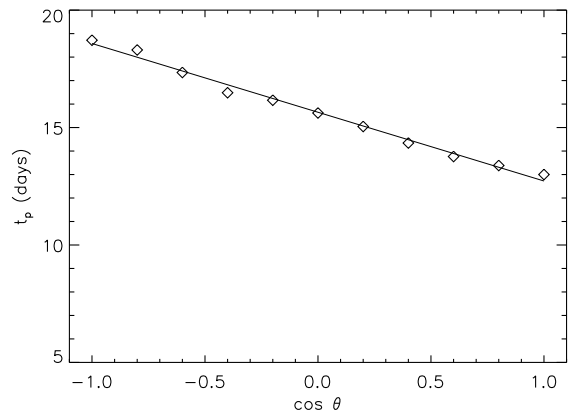, width=\columnwidth}
\caption{The time at which peak magnitude occurs ($t_{\mbox{\scriptsize p}}$)
as a function of the viewing direction ($\theta$) for light curves computed
with model~C (shown as diamonds). The solid line is the linear fit to the
results (see text).  \label{fig:toy-tpeak} }
\end{figure}

To examine the angular behaviour of the light curve properties further, one
model (model C) has been studied in greater detail. For this model, light
curves have been extracted for an additional eight viewing directions. These
directions have been selected to give a uniform grid of viewing angles in
$\cos \theta$ ($\theta$ being the angle between the line of sight
direction and the direction along which the $^{56}$Ni is displaced). Figures
\ref{fig:toy-mpeak} and \ref{fig:toy-tpeak} show the variation of $M_{\rm p}$ and
$t_{\rm p}$ with $\cos \theta$, respectively.

Figure~\ref{fig:toy-mpeak} shows that the peak magnitude follows a near-linear
dependence on $\cos \theta$. The best-fit straight line (which is drawn
in the figure) is given by:
\begin{equation}
M_{\mbox{\scriptsize p}} = -18.67 - 0.52 \cos \theta  \; \; \; .
\end{equation}
The near-linear variation with $\cos \theta$ is important since it means that
the distribution of $M_{\mbox{\scriptsize p}}$ one obtains by selecting random
lines-of-sight is close to a top-hat function; thus the model predicts
that if one were to observe it from a random viewing direction one would be
equally likely to measure any value of $M_{\mbox{\scriptsize p}}$ between about
-18.2 and -19.2 with no bias towards the median.

A similarly simple relationship is apparent in Figure~\ref{fig:toy-tpeak}. 
Again, a linear fit is an adequate description; the best-fit straight line
being given by
\begin{equation}
t_{\mbox{\scriptsize p}} = 15.7 - 2.9 \cos \theta  \; \mbox{(days)}\; \; .
\end{equation}

The absolute values for $t_{p}$ obtained from the toy models are
generally somewhat low compared to those deduced by fitting templates
to observed light curves (e.g. typically $\sim$ 19~days found in a
recent study of light curves from the Supernova Legacy Survey by 
\citealt{conley06}). This is most likely a consequence of the simple
treatment of the radiation transport which we have adopted (namely,
the use of a grey, time-independent scattering cross-section per gram), 
but it may also be, in part, a consequence of the chosen structure of the toy
models. However, a full investigation of the sensitivity of $t_{p}$ to
the assumptions made in the radiation transport and the construction of
the models goes beyond what we require for this study;
instead we focus only on the differential effects introduced by
the departures from spherical symmetry.
The relative behaviour of $t_{\mbox{\scriptsize p}}$ and
$M_{\mbox{\scriptsize p}}$ is discussed below.

\begin{table*}
\caption{Light curve properties for the toy models (A -- E). 
$\Delta v / v_{\mbox{\scriptsize max}}$ specifies the velocity offset of the
$^{56}$Ni centre-of-mass relative to the maximum velocity in the ejecta. $f$
gives the mass fraction of $^{56}$Ni in the Ni region.  $M_{\rm p}$ and
$t_{\rm p}$ are the magnitude and time at which the light curve
reaches its peak. The superscripts $\uparrow$, $\downarrow$ and $\perp$
indicate properties as viewed by a distant observer lying along the direction
in which the $^{56}$Ni is displaced relative to the SN centre-of-mass,
anti-parallel and perpendicular to this direction, respectively. 
$\Delta M_{\mbox{\scriptsize p}} = M_{\mbox{\scriptsize p}}^{\downarrow} - M_{\mbox{\scriptsize p}}^{\uparrow}$
is the range in $\Delta M_{\mbox{\scriptsize p}}$ as a function of viewing
angle for the model.}
\begin{tabular}{lcccccccccc}\\\hline
Model & $\Delta v /
v_{\mbox{\scriptsize max}}$ & $f$ & $M_{\mbox{\scriptsize p}}^{\uparrow}$ & $t_{\mbox{\scriptsize
p}}^{\uparrow}$ & $M_{\mbox{\scriptsize p}}^{\downarrow}$ & $t_{\mbox{\scriptsize
p}}^{\downarrow}$ & $M_{\mbox{\scriptsize p}}^{\perp}$ & $t_{\mbox{\scriptsize
p}}^{\perp}$ & $\Delta M_{\mbox{\scriptsize p}}$ \\ 
& & & & days & & days & & days & \\\hline
A & 0.1 & 0.45 & -19.00 & 12.0 & -18.47 & 15.2 & -18.71 & 13.5 & 0.53 \\
B & 0.1 & 0.70 & -18.95 & 13.9 & -18.43 & 16.4 & -18.69 & 15.6 & 0.52 \\
C & 0.2 & 0.70 & -19.21 & 12.7 & -18.18 & 18.1 & -18.67 & 15.7 & 1.03 \\
D & 0.2 & 0.90 & -19.19 & 13.5 & -18.14 & 18.7 & -18.64 & 16.1 & 1.05 \\
E & 0.3 & 0.90 & -19.43 & 12.0 & -17.85 & 20.3 & -18.59 & 15.9 & 1.58 \\ \hline
\end{tabular}
\label{tab:toy}
\end{table*}

\subsection{Discussion of the toy models and Arnett's Rule}
\label{sect:toy-discuss}

\begin{figure}
\epsfig{file=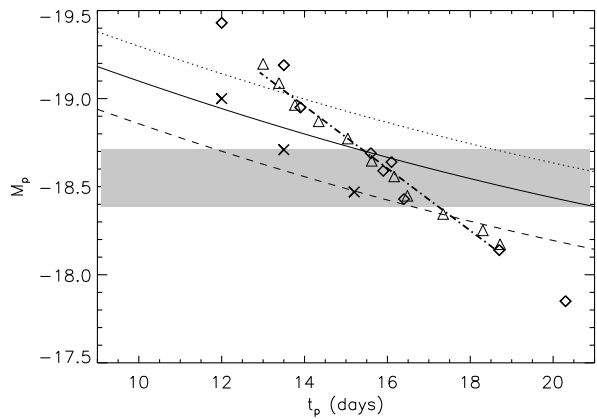, width=\columnwidth}
\caption{The peak magnitude ($M_{\mbox{\scriptsize p}}$) versus the light curve
rise time ($t_{\mbox{\scriptsize p}}$) from the toy models. There are in total
twenty-three light curves represented in the figure.  The triangles represent
the eleven light curves computed from Model C (as discussed in
Section~\ref{sect:toy-models}, these have been calculated on a uniform grid in
$\cos \theta$-space) and the heavy dot-dashed line is the description of the
Model C results obtained by combining the fits given in equations~1 and 2.  The
crosses represent the light curves computed from Model A while the
diamond symbols show those from Models B, D and E;
there are three viewing angles (parallel, perpendicular and anti-parallel to
the direction of Ni-offset) per model (the numerical values are given in
Table~1).  The three curves -- dashed, solid and dotted lines -- show the
expected Arnett relation obtained from equations 3 and 4 adopting $\alpha=0.8$,
$1.0$ and $1.2$, respectively. The shaded grey-region indicates the range of values
for $M_{\mbox{\scriptsize p}}$ predicted following
\citet{stritzinger05} for the $^{56}$Ni mass adopted in the toy models.  \label{fig:toy-arnett} }
\end{figure}

The results obtained with the toy models described above can be used to gain
some insight into how observable light curve properties behave.  In
Figure~\ref{fig:toy-arnett}, the values of $M_{\mbox{\scriptsize p}}$ are
plotted against $t_{\mbox{\scriptsize p}}$ for all the light curves summarised
in Table~\ref{tab:toy}. In addition, the grid of light curves computed to map
out the angle dependence with Model~C are also represented (eight additional
points). We emphasise that, since all the models considered adopt the same
total mass of $^{56}$Ni, the variation in $M_{\rm p}$ amongst the different points
is entirely the result of the differences in the distribution of Ni
with velocity.
Since there is no compelling reason to suppose that all observed SN~Ia synthesise the
same total mass of $^{56}$Ni -- indeed, to the contrary, 
the total mass of $^{56}$Ni is likely a dominant parameter in
distinguishing SN~Ia -- our results (Figure~\ref{fig:toy-arnett}) 
cannot be regarded as predictions
or a model for the 
directly observed distribution of SN~Ia light
curve parameters. Instead, they should be correctly interpreted in
terms of a
possible
differential effect which is sensitive to the observer line-of-sight 
and which would appear in observed distributions superimposed upon
other effects whose origins may lie in the underlying distributions of
SN~Ia explosion properties. 

As mentioned in Section~\ref{sect:toy-models}, the range of
$M_{\mbox{\scriptsize p}}$ obtained from the grid of models is very wide,
spanning more than 1.5~mag. This illustrates that the effects of a lop-sided Ni
distribution could, in principle, be very significant at the precision level
set by contemporary observational data (e.g. in a sample of SNe for which
bolometric peak magnitudes are available, as presented by
\citealt{stritzinger05}, the error estimates typically correspond to 10 -- 15
per cent in flux).  The scale of the effects is broadly comparable to those
established for other models involving departures from sphericity. For example,
\citet{hoeflich91} investigated some observational consequences of ellipsoidal
SN ejecta and showed that for quite modest departures from sphericity (axis
ratios in the range 0.9 to 1.2), one finds a viewing angle dependence of the
peak brightness on the scale of tenths of a magnitude (see also
\citealt{howell01} and \citealt{wang03} for motivation of such a model in the
context of spectropolarimetric observations). As one would expect, even greater
angular variations -- comparable to those obtained here -- can be obtained from
ellipsoidal models with more extreme axis ratios \citep{sim07}. Similar scales
of light curve angular dependence (up to $\sim 0.25$~mag) were also 
found by \citet{kasen04}
in a study of a SN model with a geometric ``hole'' in the ejecta
and by \citet{kasen06c} in a very recent study of a 2D
GCD model.

We note that the angular variation is so large that one can likely rule out the
more extreme models (models C, D and E) as being representative of typical SNe
conditions -- this follows from the observed tightness of the correlation in
the plot of bolometric flux versus recession velocity for nearby SNe~Ia (the
``Hubble diagram''; see e.g. \citealt{stritzinger05}). However, these models do
show that it is possible to arrange extreme geometries that could produce
substantial deviations from the mean and provide one possible explanation for
some apparently anomalous events \citep{hillebrandt07}.

Since both $M_{\rm p}$ and $t_{\rm p}$ are simple functions of the viewing
angle (equations 1 and 2), it follows that there is an equally simple
relationship between these two quantities for a given model, as can be seen in
Figure~\ref{fig:toy-arnett}. For the models with moderate to large values of $f$
(Models B -- E), the points are all fairly close to following the same line in
the $M_{\rm p}$-$t_{\rm p}$ plane. Thus, from these models alone one might hope
that combined measurements of both $M_{\rm p}$ and $t_{\rm p}$ could be used to
extract the true luminosity despite the effect of viewing angle. However, the
points obtained with model~A lie significantly below this line, showing that
the relationship is not universal.

A point of particular interest relates to the interpretation of Arnett's Rule
\citep{arnett82} which is commonly used to estimate Ni masses in studies of
SNe~Ia light curves (e.g. \citealt{arnett85}; \citealt{branch92};
\citealt{stritzinger05}; {\citealt{stritzinger06a};
\citealt{stritzinger06b}}; \citealt{howell06}).  Arnett's Rule states that at
maximum light, the luminosity roughly balances the instantaneous rate of energy
generation in the SN. Adopting the convenient notation of
\citet{branch92} this relation is stated as
\begin{equation}
L^{\mbox{\scriptsize p}} = \alpha R(t_{\mbox{\scriptsize p}}) M_{
\mbox{\scriptsize Ni}}
\end{equation}
where $L^{\mbox{\scriptsize p}}$ is the peak luminosity,
$R(t_{\mbox{\scriptsize p}})$ is the rate of energy generation at the time of
maximum light ($t_{\mbox{\scriptsize p}}$) and $M_{ \mbox{\scriptsize Ni}}$ is
the total mass of $^{56}$Ni. The proportionality factor, $\alpha$, is expected
to be of order unity. $R(t)$ is given by
\begin{equation}
R(t) = [6.41 \; e^{-t/8.8} + 1.34 \; e^{-t/113.7}] 10^{43}\; \; \mbox{ergs s$^{-1}$ M$_{\odot}^{-1}$}
\end{equation}
where $t$ is measured in days. The numerical coefficients have been computed
using the lifetimes and decay energies from \citet{ambwani88}.

It is well-known from several previous studies that Arnett's Rule can usually
be used to derive a reasonable estimate for $M_{\mbox{\scriptsize Ni}}$ from
the light curve. However, it is not strictly obeyed, particularly if there is
Ni present in the outermost regions of the SN ejecta, as discussed by
\citet{pinto00a}.  Since Arnett's Rule relates only global quantities, for any
class of SN model in which a viewing angle dependence of the flux appears
(including, but not limited to, the off-centre models discussed here), an
attempt to measure the nickel-mass via Arnett's Rule must introduce a
systematic error which depends upon the direction of observation.  In this
section, we will use the toy models to investigate how this systematic would
behave if real SNe (or at least a subset of them) were to harbour lop-sided
distributions of $^{56}$Ni.

First we consider the case where a Ni mass is deduced by the application of
Arnett's Rule (with a chosen value, or range of values, for $\alpha$) to direct
measurements of the observed peak magnitude and time of maximum light. To
elucidate this case, we over-plot in Figure~\ref{fig:toy-arnett} the curves
obtained by combining equations~3 and 4 for three different values of $\alpha$
(0.8, 1.0, 1.2). It comes as no surprise that Arnett's Rule with a fixed value
of $\alpha$ does not describe the co-dependence of $M_{\rm p}$ and 
$t_{\rm p}$ 
obtained from the 
models -- the Rule is concerned with relating global properties of
different SNe, not the detailed angular dependence within single objects.
However, the correlation of $M_{\rm p}$ and $t_{\rm p}$ deriving from Arnett's Rule
{\it is} in the same sense as that obtained from the models. This helps to
suppress the systematic error one would introduce with an unknown viewing
angle; for the more moderate models (A and B), the systematic error introduced
would be only around 0.15 to 0.2~mag. In exceptional cases, however, a much
larger systematic error of up to about 0.5~mag is possible for the adopted Ni
mass. 

Secondly, we consider the case where reliable measurement of $t_{\rm p}$ has
not been possible so that an estimate of $M_{\mbox{\scriptsize Ni}}$ must be
obtained from $M_{\rm p}$ alone.  Adopting the relationship discussed by
\cite{stritzinger05} (their equation~7 which is derived from Arnett's Rule
and an empirically motivated assumption for the value of $t_{\rm p}$), leads
to an expected value of $M_{\rm p} = -18.56 \pm 0.16$~mag for
$M_{\mbox{\scriptsize Ni}} = 0.4$~M$_{\odot}$. This range of $M_{\rm p}$ is
indicated in Figure~\ref{fig:toy-arnett} by the grey shaded region; thus all
the points which fall within this band are consistent with the
\cite{stritzinger05} relationship.  For all the models considered, a
significant fraction of the possible viewing directions lead to light curves
which lie outside the range. Thus, if the range of light curve properties
produced by the models were representative of those from real SNe, using
that relationship between $M_{\rm p}$ and $M_{\mbox{\scriptsize Ni}}$ in the
analysis of a sample of observed light curves could lead one to infer a wider
spread of nickel masses than required.

The toy models have demonstrated that it is possible to construct very
simplistic scenarios in which a lop-sided distribution of $^{56}$Ni can
introduce significant angular dependence in the emergent radiation. The scale
of the variation is comparable to that introduced by other types of departure
from spherical symmetry (\citealt{hoeflich91}; \citealt{kasen04}) and can
be comparable to, or larger than, typical observational uncertainties. 
However, the toy models are very simplistic and are derived with no reference
to realistic explosion physics. Therefore, they merely illustrate possible
effects and one must appeal to 3D explosion models to judge whether such
effects are likely to have a part to play in reality. The remainder of this
paper is concerned with the analysis of results from one such model.

\section{A 3D explosion model}
\label{sect:real_model}

\subsection{The model}
\label{sect:3d}

\begin{figure*}
\epsfig{file=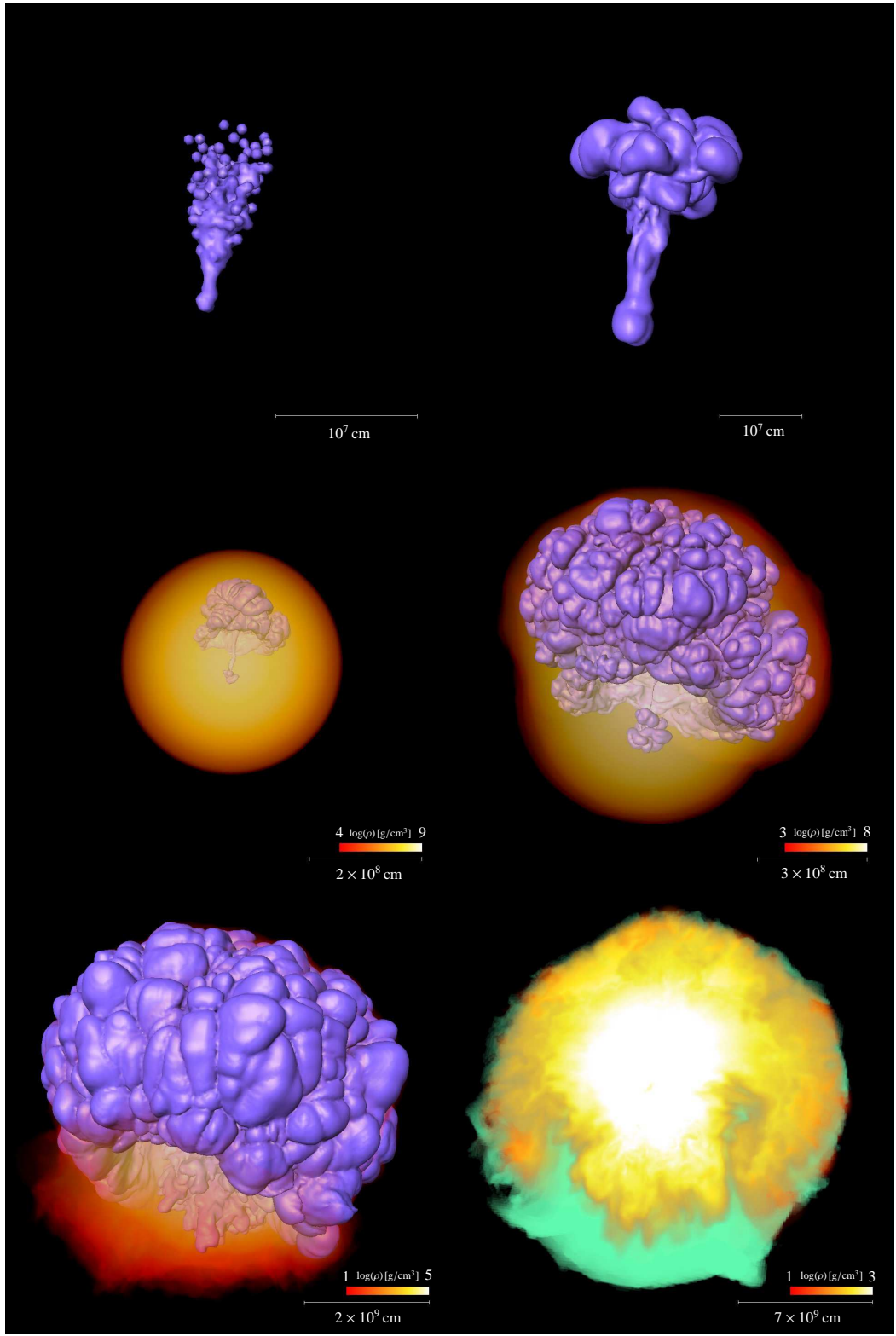, height=0.87\textheight}
\caption{Evolution of the 3D simulation 3T2d200. The initial flame
  configuration is shown in the upper left panel; 
  the blue isosurface corresponds to the zero level set 
  associated with the thermonuclear flame.
  The following four panels ($t = 0.2\,\mathrm{s}$, $t = 0.4 \, \mathrm{s}$,
  $t = 1.0 \, \mathrm{s}$, and $t = 3.0 \, \mathrm{s}$;
  left to right, top to bottom) illustrate the evolution of the
  model. Again,
  the blue isosurface corresponds to the zero level set
  associated with the
  flame where burning is active and gives an approximate interface between fuel
  and ashes where the densities are below the burning threshold 
  ($10^7 \, \mathrm{g}\,\mathrm{cm}^{-3}$). The star is indicated in some of
  the figures by the volume-rendering of the density (red through
  yellow to white). The lower right image
  illustrates the composition of the ejecta (the front hemisphere is cut away)
  at the end of the simulation ($t = 10.0 \, \mathrm{s}$). Unburned material
  (carbon and oxygen) is shown in green and the 
  white/yellow/orange volume-rendering
  corresponds to the density of the iron group elements.  \label{fig:td} }
\end{figure*}

The 3D explosion simulation 3T2d200 described by \citet{roepke07} has been
used as the basis for a model to explore the effects of an off-centre 
$^{56}$Ni distribution
in a realistic case. This simulation followed the flame evolution when ignited
in a lop-sided teardrop-like shape (upper left panel of Figure~\ref{fig:td}). 
Such an ignition configuration is motivated by recent studies of the
pre-ignition convective burning phase \citep{kuhlen06}, which suggest that the
flow pattern is dominated by a dipole at this stage. A consequence could be a
lop-sided ignition as adopted here, where the majority of ignition kernels are
located on one side of the star but with a slight over-shooting across 
the centre. In this case, the flame propagates in both directions, subject to
buoyancy instabilities and accelerated by turbulence. One side of the flame
structure, however, remains dominant (see the top left panel in 
Figure~\ref{fig:td}). Although the ash bubble on this side of the 
star starts to
sweep around the core (similar to what has been described by
\citealt{plewa04}), it does not collide on the far side (see the panels in the
second row of Figure~\ref{fig:td}) because the 
energy generated by nuclear burning
is sufficient to gravitationally unbind the star and expand it before a
collision occurs. The lower left panel of Figure~\ref{fig:td} 
shows the end-stage
of this evolution; subsequently the expansion is approaching homology. The
composition of the ejecta is illustrated in the lower right panel of
Figure~\ref{fig:td}, showing a pronounced asymmetric bubble of iron-group
elements. The simulation was carried out on a 512$^3$ cells moving Cartesian
computational grid \citep{roepke05b}.

The model used here for radiation transport adopts a uniform Cartesian
grid with 128$^3$ cells. The total mass density and 
the mass fraction of Fe-group elements
in each cell was obtained directly from the hydrodynamic simulation
by combining sets of 4$^3$ cells from the original 512$^3$ grid. The cell
properties were extracted from the latest time to which the hydrodynamics
calculation was followed ($\sim$10s). For times beyond this point, including
the entirety of the time for which the radiative transfer is followed, the
ejecta are assumed to be in homologous expansion.

Since the hydrodynamical results did not provide detailed information on the
composition of the ash material, it has been assumed that initially all the
Fe-group mass was composed of $^{56}$Ni; this gives a total of 0.448~M$_{\odot}$
of $^{56}$Ni.  Since this investigation is concerned with the viewing-direction
dependence of the light curve, this assumption will not significantly affect
our conclusions -- a lower mass fraction of $^{56}$Ni in the ash would merely
reduce the total luminosity; it would only affect the angle-dependence if the
mass fraction were to vary significantly with position; whether or not this
could be the case must await more detailed nucleosynthesis modelling in 3D.

\subsection{Treatment of opacity}

In the calculations using the 3T2d200 model, a slightly more sophisticated
treatment of {\sc uvoir} opacity is employed 
than the uniform grey cross-section adopted
for the toy models.  Following \citep{mazzali06}, an opacity which depends on
composition is adopted, the particular form used here being
\begin{equation}
\sigma = 0.26 \; (0.9 X_{\mbox{\scriptsize Fe-grp}} + 0.1) \; \mbox{cm$^{2}$~g$^{-1}$}
\end{equation}
where $X_{\mbox{\scriptsize Fe-grp}}$ is the mass-fraction of iron-group
elements, which varies throughout the model. This simple parameterisation takes
advantage of the basic compositional information available directly from the
hydrodynamics calculation (namely the mass-fraction of heavy elements) to
attempt to account for the dominance of the iron-group elements in providing
opacity.

As has been discussed elsewhere (e.g. see \citealt{pauldrach96} 
and \citealt{pinto00b} for in-depth
discussion or \citealt{sim07} for comments in the context of the code used
here), the {\sc uvoir}-opacity in SN ejecta is mostly dominated by spectral
lines. Given this, the grey-approximation adopted here is not valid in detail
-- in particular, it cannot account for the process of photon redistribution in
frequency-space which allows photons to escape the ejecta by down-scattering
into frequency regimes in which the density of spectral lines is low. This
process effectively limits the maximum trapping any photon can undergo, an
effect which is most relevant at the earliest times. For this reason, the
grey-approximation may overestimate the dependence of radiative transfer
effects on geometry. Nevertheless, the grey approximation is retained here
since it dramatically reduces the computational demands of the radiation
transfer calculations compared to a full non-LTE calculation.  Despite this
approximation, the computations can be expected to correctly identify the sense
in which purely geometrical effects will act and to give a simple estimate, or
at least a reliable limit, for their scale.

Of course, fully 3D, non-LTE, non-grey calculations are necessary to
quantify the effects rigorously
-- however, to utilise this level of sophistication in the radiative
transfer fully, 
one would also require more complete information on the detailed
nucleosynthesis products and the ultimate evolution toward homology than is
presently available for the 3T2d200 model. Fortunately, as discussed by
\citet{lucy05}, the Monte Carlo radiative transfer method is well-suited to
incorporating more advanced treatments of microphysics and thus there is great
promise for more detailed treatments when merited (see, e.g.
\citealt{kasen06a}, \citealt{kasen06c}).

\subsection{Results obtained with the explosion model}
\label{sect:3d-discuss}

\begin{figure}
\epsfig{file=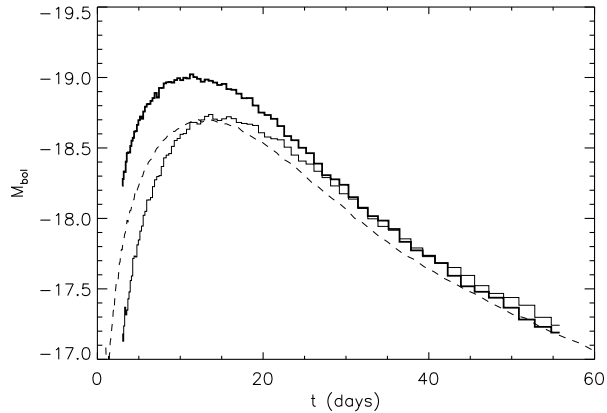, width=\columnwidth}
\caption{Computed light curves from the 3T2d200 model \citep{roepke07}.  The
  dashed line shows the angle-averaged light curve. The heavily and lightly drawn
  histograms show, respectively, the light curve as seen by distant observers who
  lie along the direction-vectors {\boldmath$\hat{n}$} and {\boldmath $-\hat{n}$}. 
  {\boldmath$\hat{n}$} is the unit-vector which points from the centre-of-mass of
  the ejecta to the centre-of-mass of ash material.  \label{fig:3d-lcs}
}
\end{figure}

We begin by presenting and describing the light curve properties obtained from
the 3T2d200 model.  Figure~\ref{fig:3d-lcs} shows light curves obtained for the
special viewing directions parallel and anti-parallel to the vector connecting
the centre-of-mass of the ejecta and the centre-of-mass of the $^{56}$Ni;
throughout the discussion we denote this vector {\boldmath $n$} -- note that
{\boldmath $n$} is nearly, but not quite, parallel to {\boldmath $\hat{z}$}.

We see from the figure that there are differences in the light curve properties
as seen from these two directions -- in agreement with the results from the toy
models, the scale of the effect is significant compared to typical
observational accuracy ($\leq 0.1$~mag).  Also in accordance with the results
obtained in Section~\ref{sect:toy-models}, the light curve is brightest and
rises to maximum fastest when viewed down {\boldmath $n$}. In contrast to the
toy-model results, however, the anti-parallel direction does not show the
dimmest light curve for the 3D model -- this is a consequence of the more
complex distribution of ash material which determines both the location of
energy generation ($^{56}$Ni) and the distribution of opacity (via the
dependence on $X_{\mbox{\scriptsize Fe-grp}}$ 
in equation~5). As might be anticipated from
Figure~\ref{fig:td} (lower panels), the dimmest light curves are actually
seen by observers whose line-of-sight is nearly perpendicular to 
{\boldmath $n$}
(see below).

\begin{figure}
\epsfig{file=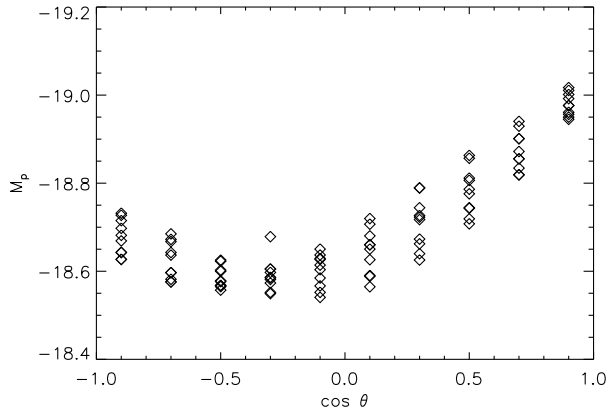, width=\columnwidth}
\caption{The peak magnitude as a function of the spherical polar angle
  $\theta$ as obtained with the 3T2d200 model. For each value of $\theta$ there
  are 10 points, each representing a different azimuthal angle, $\phi$.
  \label{fig:3d-mpeak}
}
\end{figure}

While the light curves shown in Figure~\ref{fig:3d-lcs} indicate that it is
possible to see significant differences in the peak magnitude as a function of
direction, it is important to consider how probable a randomly aligned observer
is to see a given range of light curve parameters. To this end, a grid of 100
viewing directions (evenly spaced in solid angle), has been employed to map out
the angular distribution of the light curve properties.  In
Figure~\ref{fig:3d-mpeak} the peak magnitudes are shown for this grid of
directions as a function of the polar angle, $\theta$, ($\theta$ being 
the angle
between the $z$-axis of the model and the observer's line-of-sight).  
There is clearly a net trend in the variation of $M_{\rm p}$ with
$\theta$; for large values of $\cos \theta$ the mean trend is very similar to
that found in the toy models. But, as noted above, owing to the complex
distribution of ash material, a much weaker trend is revealed in the region
where $\cos \theta < 0$.

\begin{figure}
\epsfig{file=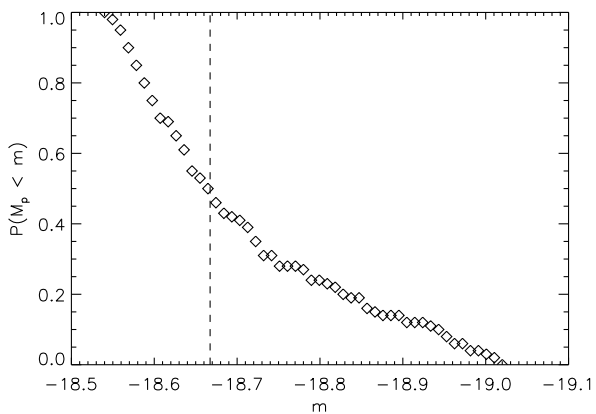, width=\columnwidth}
\caption{The peak-magnitude cumulative probability distribution function -- the
  probability that the obtained value of $M_{\rm p}$ is less than $m$ -- 
  as obtained from the 3T2d200 model.  
  The median value is -18.67 (dashed line).
\label{fig:p-dist}}
\end{figure}

The grid of light curves as a function of viewing directions can be used to
determine the probability distribution for the peak magnitude.  This has been
done and the resulting cumulative probability distribution is shown in
Figure~\ref{fig:p-dist}.

Although more complex in detail, the probability distribution function (pdf)
obtained shares most of the important properties suggested by the toy
models; namely, it is fairly wide (spanning $\sim$ 0.5 mag) and the probability
is not strongly concentrated around the median (for example, fully one quarter
of the directions give peak magnitudes that are brighter than the median by at
least 0.1 mag).  An important difference from the pdf implied by the toy models
is a distinct asymmetry (as could be anticipated from
Figure~\ref{fig:3d-mpeak}) -- there is a probability tail extending to brighter
magnitudes meaning that one may see larger differences on the bright side than
on the dim side of the median.  The probability distribution for $M_{\rm p}$
will be used in Section~\ref{sect:hubble} to quantify the relevance of this
work to the interpretation of the local Hubble diagram. 

\begin{figure}
\epsfig{file=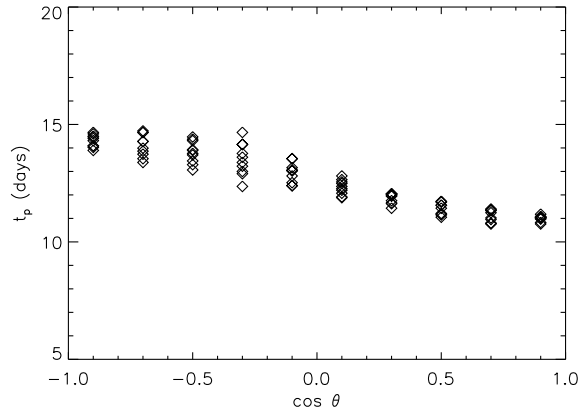, width=\columnwidth}
\caption{As for Figure~\ref{fig:3d-mpeak} but showing the time at which maximum light occurs.
\label{fig:3d-tpeak}
}
\end{figure}

For completeness, Figure~\ref{fig:3d-tpeak} shows the range of light-curve rise
times with viewing direction. As with the toys models, it can be seen that
there is significant diversity in the rise time and that it tends to be
shortest when viewing down {\boldmath $n$} (i.e. close to $\theta = 0$) and
longest when viewing closer to {\boldmath $-n$} (i.e. $\theta \sim \pi$).  In
the next section, we examine the relationship between rise time and peak
magnitude in some detail.

\section{Implications of the 3D calculations}
\label{sect:implic}

We now consider some ramifications of the results obtained from the 3D
model that has been discussed in Section~\ref{sect:real_model}.

\subsection{Arnett's Rule and observational estimates of Ni masses}

In Section~\ref{sect:toy-discuss} we used the toy models to investigate the
possible systematic effects of an off-centre explosion on the inference of Ni
masses. Here, we extend that discussion using the results obtained with the
3T2d200 explosion model.

\begin{figure}
\epsfig{file=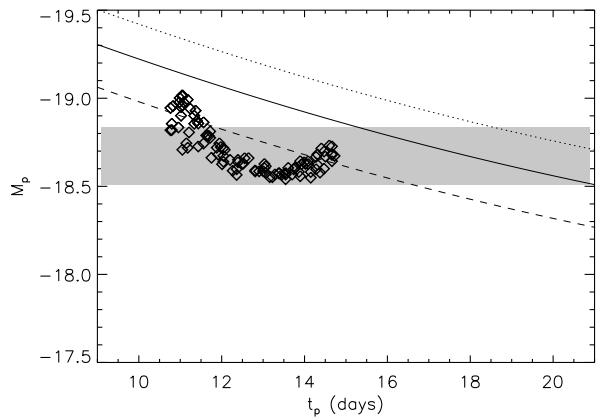, width=\columnwidth}
\caption{ The peak magnitude ($M_{\mbox{\scriptsize p}}$) versus the light
  curve rise time ($t_{\mbox{\scriptsize p}}$) from the 3T2d200 model
  (R\"{o}pke et al. 2006).  The opacity depends on composition via equation~5. 
  The three curves -- dashed, solid and dotted lines -- show the expected
  Arnett relation obtained from equations 3 and 4 adopting $\alpha=0.8$, $1.0$
  and $1.2$, respectively. The shaded grey-region indicates the range of values
  for $M_{\mbox{\scriptsize p}}$ predicted following \citet{stritzinger05} for
  a $^{56}$Ni mass of 0.488~M$_{\odot}$.  \label{fig:3d-arnett} }
\end{figure}

Figure~\ref{fig:3d-arnett} shows the peak magnitude versus the time of maximum
light for the set of 100 light curves obtained for different viewing directions
in the 3T2d200 model.  As in Figure~\ref{fig:toy-arnett}, curves are
over-plotted to indicate the trends predicted by equations~3 and 4.  Similarly,
the grey band indicates the region in which the computed values of 
$M_{\rm p}$ are
consistent with the relationship presented by \citep{stritzinger05}.

The figure reveals several interesting phenomena. Amongst the highest
luminosity and fastest rise-time light curves there remains a trend which is
reminiscent of what was suggested by the toy models
(Figure~\ref{fig:toy-arnett}). However, this trend does not continue through
the light curves with longer rise times; instead, the trend first flattens and
then turns over -- a result of the non-linear angular dependence of
$M_{\rm p}$ as shown in Figure~\ref{fig:3d-mpeak}.  This has the consequence
that, as the viewing angle is varied, the light curve properties stay fairly
close to those inferred via Arnett's Rule (with $\alpha \sim 0.8$) but with
systematic variation of around $\sim 0.2$~mag ($\sim 20$ per cent in luminosity).

The value for the peak bolometric magnitude obtained using the model
$^{56}$Ni mass
and following \citet{stritzinger05} is very close to the median of the peak
magnitude obtained amongst the different viewing angles. The error range which
they suggest -- based on an estimated range of possible light curve rise times
-- happens to be rather close to encompassing the range of peak magnitudes
obtained from the model. However, there are a moderate number of model light
curves which lie outside this range, all of which are brighter than
the
\citet{stritzinger05}
relation would anticipate.

Thus we conclude that the calculations involving the fully 3D model support the
primary result obtained with the toy models, namely that observationally
significant systematic errors can be introduced via viewing angle effects in
off-centre explosions. The particular model used here has a maximum spread in
$M_{\rm p}$ of about 0.5~mag, which is comparable to that obtained with toy
model A. However, the particularly simple, linear dependence of $M_{\rm p}$
with $\cos \theta$ (see Figure~\ref{fig:toy-mpeak}) is not preserved in the 3D
model. 

\subsection{Scatter in the local Hubble diagram}
\label{sect:hubble}

In this section we investigate what effect the predicted viewing angle
dependence would have on the observed relationship between peak magnitude and
recession velocity for SNe~Ia in the Hubble flow.
    
\begin{figure}
\epsfig{file=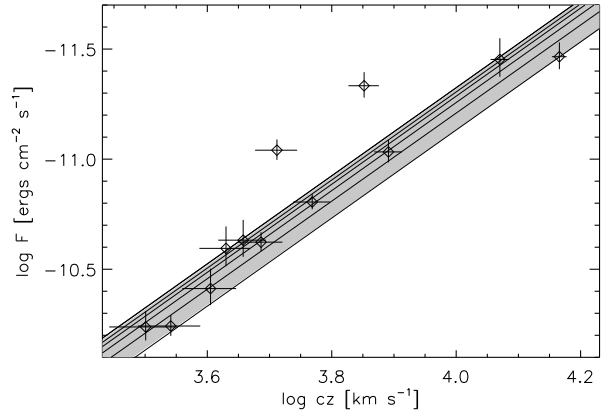, width=\columnwidth}
  \caption{ The peak {\sc uvoir} flux ($F$) versus recession velocity ($cz$)
  for SNe~Ia in the local Universe.  The grey shaded area indicates the region
  which would be populated by a multitude of 3T2d200 SNe in the
  Hubble flow having random orientations relative to the observer
  line-of-sight.  The solid lines are drawn to delineate equal intervals of the
  underlying probability distribution for $M_{\rm p}$ (they show 20 per cent
  intervals).  The twelve data points which are over-plotted are those for
  the SNe~Ia discussed by \citet{stritzinger05}.  \label{fig:3d-hubble} }
\end{figure}

Assuming that the properties of the 3T2d200 model (Figure~\ref{fig:3d-mpeak})
are representative for a population of SNe~Ia, 
the viewing-angle dependence of the
bolometric light curve would lead to a scatter about the mean relationship
between peak flux and recession velocity.  To quantify this effect, we have
used the probability distribution of the peak magnitude
(Figure~\ref{fig:p-dist}) to map the region of the local Hubble diagram that
would be occupied by a hypothetical population of SNe~Ia 
which are in the Hubble
flow and which have properties as predicted by the 3T2d200 model.  This region
is indicated by the shaded area in Figure~\ref{fig:3d-hubble}. Over-plotted are
lines which show contours of the probability distribution; these contours are
slightly concentrated toward the fainter side owing to the asymmetry of the
probability distribution (see Section~\ref{sect:3d-discuss}).

In constructing the Hubble diagram (via equation~10 of Stritzinger \&
Leibundgut 2005) we have adopted a value for the Hubble parameter, $H_{0} = 85$~km~s$^{-1}$~Mpc$^{-1}$. 
This value is the one that \citet{stritzinger05} obtain from their
analysis using a typical $^{56}$Ni mass of 0.42~M$_{\odot}$; as they discuss,
this value for $H_{0}$ is unusually large and most likely a result of a net
under-prediction for the amount of $^{56}$Ni produced in the pure deflagration
explosion models that they discuss.  Although this is clearly an important
issue, it is not of relevance to our discussion of scatter in the Hubble
diagram since the effect of changing the adopted value of $H_{0}$ amounts to
only a fixed offset in the ordinate of Figure~\ref{fig:3d-hubble}.

For comparison, the flux and recession velocities for the sample of twelve
local SNe~Ia 
which was assembled and used by \citet{stritzinger05} are also shown
in the figure. Although relatively small compared to data sets
obtained from the large scale supernova surveys, 
the \citet{stritzinger05} sample is ideally-suited for  
comparison with our results since the objects they include have been
selected on the basis of having sufficiently large wavelength coverage 
that the
bolometric peak magnitude is well-determined observationally; this is
important since our calculations do not distinguish between different
photometric bands. Also, the fluxes tabulated by
\citet{stritzinger05} have not been normalised or modified using any
empirically motivated stretch factors or light curve shape
correlations. Since we do not
compute band-limited light curves, we cannot use the
standard empirical correlations quantitatively (e.g. the \citealt{phillips93} relation)
to renormalise our calculations. Therefore, for quantitative
comparison of like with like, we require the distribution of observed
bolometric fluxes corrected only for observational effects such as
reddening. 

Figure~\ref{fig:3d-hubble} clearly demonstrates 
two important points which we now
discuss in turn.
First, it shows that the viewing angle variation predicted by the 3T2d200 model
does not substantially exceed the scale of scatter in the observed Hubble
diagram. Therefore, one cannot rule out the 3T2d200 model as representative of
a typical SN~Ia by arguing that its asphericity is inconsistent with the
observed narrowness of the Hubble-diagram correlation. This is in contrast to
the more extreme toy models (see Section~\ref{sect:toy-discuss}) which could
be excluded by this argument directly.

Secondly, it can also be seen that the angular variations in the peak magnitude
are on the correct scale to account for the observed dispersion about the mean
relation. Thus it is probable that orientation effects in aspherical
SNe~Ia could
contribute significantly to the dispersion in the Hubble diagram. Indeed we
conclude that, excluding the two apparently sub-luminous objects (SN1992bo and
SN1993H), the width of the distribution of bolometric magnitudes in the sample
of well-observed objects selected by \citet{stritzinger05} can be 
explained fully by the viewing angle 
effects predicted from the 3T2d200 model without
the need to invoke any other systematic or random uncertainty.

If the scatter found in the bolometric model light curves translates into a
similar scatter in band limited light curves, this effect would have important
implications for the use of SNe~Ia as 
standard candles: as a real physical effect
which is dependent on the explosion mechanism, it must be understood and cannot
be eliminated as a source of dispersion by improved observational statistics or
measurements alone. In addition, the finding that the probability distribution
is not symmetric around the mean (cf. Figure \ref{fig:p-dist}) could
potentially introduce an additional bias in the interpretation of observed
SN samples -- if a significant fraction of the observed SN~Ia sample has
properties similar to the ones found in this model, the probability to find
objects that are fainter than the mean may be somewhat larger than finding ones
brighter than the mean.

The commonly used methods to correct the peak magnitude based on the shape of
the light curve in certain filter bands (e.g., \citealt{phillips93},
\citealt{riess96}) are not expected to account for the orientation
effect fully. 
The calibration methods address a scatter in the band-limited peak flux
which is most likely introduced by different degrees of nuclear burning and
mixing which affect the emerging radiation in specific filter bands
(\citealt{hoeflich95}; \citealt{pinto01a}; \citealt{mazzali01};
\citealt{kasen06b}) through the resulting variation in the composition and
thermal structure of the ejecta. Because it is unlikely that the orientation
effects follow the same correlation, it is expected that these two sources of
scatter will be superimposed in the distribution of observed peak magnitudes in
particular photometric bands.

It is conceivable that a relation between observables can be found that allows
a correction following the spirit of the Phillips relation. However, the
calculations presented here are not suitable to address this question because
band-limited light curves, and likely also simulated spectra for the 3D model,
are required to consider this point in detail.  Knowing how the orientation
effect affects the band limited light curves would also enable us to analyse
the much larger data sets that are available for SNe~Ia observed in specific
wavelength bands to estimate the significance of asphericity in the observed
objects.

\section{Summary}
\label{sect:summ}

Motivated by recent explosion models involving off-centre ignition in
SNe~Ia, we
have investigated the possible range of effects of a lop-sided distribution of
nuclear ash on bolometric light curves for such objects. 

By studying results from both a grid of simple toy models and one real
explosion model, we find that off-centre distributions of material which has
undergone nuclear burning are likely to leave detectable imprints on observed
light curves. An angular dependence of the light curve peak brightness is
introduced -- based on the models considered, the scale of this effect can
be readily $\sim 0.2$~mag and conceivably much greater under extreme
circumstances.

This effect is large enough to have ramifications for the interpretation of the
diversity in observed SN~Ia properties and we have shown that the explosion
model which we considered \citep{roepke07} already predicts sufficient
viewing-angle sensitivity to explain much of the typical scatter in a sample of
objects with well-observed bolometric properties \citep{stritzinger05}.

In view of the potential observational significance, further study of this
class of explosion model is required. Investigation of viewing angle effects
across a more diverse range of observable quantities is warranted (i.e.
photometric band-limited light curves and, ultimately, spectra) to establish
whether and by what means such effects may be observationally distinguished
from other factors which determine light curve properties (e.g. the
total $^{56}$Ni
mass). Furthermore, a wider investigation of the possible extent and frequency
of off-centre ignition in explosion models is needed so that the likelihood for
occurrence of such events -- and therefore the potentially important
systematics they may introduce to statistical analysis of the SN~Ia sample --
can be quantified.

\section*{Acknowledgments}

We thank an anonymous referee for insightful comments on the manuscript.
This work was supported in part by
the Deutsche Forschungsgemeinschaft through the Transregional
Collaborative Research Centre TRR 33 ``The Dark Universe''.

\bibliographystyle{mn2e}
\bibliography{snoc}

\label{lastpage}

\end{document}